\documentclass{emulateapj}

\begin{document}
\title{Radiation Mechanism and Jet Composition of Gamma-Ray Bursts and GeV-TeV selected Radio Loud Active Galactic Nuclei}
\author{Jin Zhang\altaffilmark{1}, En-Wei Liang\altaffilmark{2,1}, Xiao-Na Sun\altaffilmark{2}, Bing Zhang\altaffilmark{3}, Ye Lu\altaffilmark{1}, Shuang-Nan Zhang\altaffilmark{1,4}}
\altaffiltext{1}{National Astronomical Observatories, Chinese Academy of Sciences, Beijing 100012, China}\altaffiltext{2}{Department of Physics and GXU-NAOC Center for Astrophysics
and Space Sciences, Guangxi University, Nanning 530004, China; lew@gxu.edu.cn} \altaffiltext{3}{Department of Physics and Astronomy, University of Nevada, Las Vegas, NV 89154, USA} \altaffiltext{4}{Key Laboratory of Particle
Astrophysics, Institute of High Energy Physics, Chinese Academy of Sciences, Beijing 100049,
China}

\begin{abstract}
Gamma-ray bursts (GRBs) and GeV-TeV selected radio loud Active Galactic Nuclei (AGNs)
are compared based on our systematic modeling of the observed spectral energy distributions of a sample of AGNs with a single-zone leptonic model. We show that the correlation between the jet power ($P_{\rm jet}$) and the prompt gamma-ray luminosity ($L_{\rm jet}$) of GRBs is consistent, within the uncertainties, with the correlation between jet power and the synchrotron peak luminosity ($L_{\rm s, jet}$) of flat spectrum radio quasars (FSRQs). Their radiation efficiencies ($\varepsilon$) are also comparable ($>10\%$ for most sources), which increase with the bolometric jet luminosity ($L_{\rm bol,jet}$) for FSRQs and with the $L_{\rm jet}$ for GRBs with similar power-law indices. BL Lacs do not follow the $P_{\rm jet}-L_{\rm s, jet}$ relation of FSRQs. They have lower $\varepsilon$ and $L_{\rm bol, jet}$ values than FSRQs, and a tentative $L_{\rm bol, jet}-\varepsilon$ relation is also found, with a power-law index being different from that of the FSRQs. The magnetization parameters ($\sigma$) of FSRQs are averagely larger than that of BL Lacs. They are anti-correlated with $\varepsilon$ for the FSRQs, but positive correlated with $\varepsilon$ for the BL Lacs. GeV Narrow-line Seyfert 1 galaxies potentially share similar properties with FSRQs. Based on the analogy between GRBs and FSRQs, we suggest that the prompt gamma-ray emission of GRBs is likely produced by synchrotron process in a magnetized jet with high radiation efficiency, similar to FSRQs. The jets of BL Lacs, on the other hand, are less efficient and are likely more matter dominated.
\end{abstract}


\section{Introduction}           
\label{sect:intro}
The emission of gamma-ray bursts (GRBs) and radio loud active galactic nuclei (AGNs) is generally believed to be produced in relativistic jets powered by their central black holes (BHs). GRBs are the most luminous transients in the universe. Their progenitors are thought to be either core collapses of massive stars (Woosley 1993; Paczy\'{n}ski 1998) or mergers of two compact objects (Eichler et al. 1989). Such a catastrophic event likely gives birth to a new-born stellar-mass BH, which accretes mass from a torus. An ultra-relativistic jet is launched from the central engine, which dissipates its kinetic or magnetic energy at a large radius to power the gamma-rays observed. Radio-loud AGNs are believed to be powered by a super-massive rotating BH, which launches a mildly relativistic jet (Urry \& Padovani 1995). Most confirmed extragalactic GeV-TeV sources are blazars, a sub-sample of radio-loud AGNs. They are divided into flat spectrum radio quasars (FSRQs) and BL Lac objects (BL Lacs), based to whether or not strong emission line features are observed.

The broad band spectral energy distributions (SEDs) of blazars usually show two bumps, which are generally explained as synchrotron radiation and inverse Compton (IC) scattering of a same population of relativistic electrons, respectively. The seed photons for the IC process can be from the synchrotron radiation itself (i.e., the synchrotron self-Compton [SSC] model, Maraschi et al. 1992; Ghisellini et al. 1996; Zhang et al. 2012b), or from an external radiation field (e.g., the external Compton [EC] model, Dermer et al. 1992). Narrow-Line Seyfert 1 Galaxies (NLS1s) were also identified as a new class of GeV gamma-ray AGNs by the Fermi/LAT (Abdo et al. 2009a). Their broad band SEDs are similar to that of FSRQs, which are also well explained by this synchrotron+IC leptonic jet model (Abdo et al. 2009a).

In contrast to blazars whose radiation mechanisms (especially for the low-frequency component) are well understood, the radiation mechanism of GRB prompt gamma-ray emission is still not identified. This prompt emission spectrum is usually well described by a smooth broken power-law function called the ``Band function'' (Band et al. 1993), with a spectral peak typically in the range of 100s of keV. The origin of this spectrum is under intense debate. Although synchrotron emission has been proposed as the leading model (e.g., M\'esz\'aros et al. 1994; Wang et al. 2009; Daigne et al. 2011; Zhang \& Yan 2011), other mechanisms including Comptonization of quasi-thermal photons from the fireball photosphere (e.g., Rees \& M\'esz\'aros 2005; Pe'er et al. 2006; Giannios 2008; Beloborodov 2010) and synchrotron self-Compton (e.g., Racusin et al. 2008; Kumar \& Panaitescu 2008) have been also suggested. Even though photosphere emission is believed to dominate the spectrum in a small fraction of GRBs such as GRB 090902B (Abdo et al. 2009b; Ryde et al. 2010; Zhang et al. 2011) and probably GRB 090510 (Ackermann et al. 2010), the Band component in most GRBs may not be the modified photosphere emission. In particular, Zhang et al. (2012a) show that the peak energy of the Band component of GRB 110721A is beyond the ``death line'' in the $L_{\rm iso}-E_{\rm p}$ diagram for the photosphere models, where $L_{\rm iso}$ is the isotropic prompt gamma-ray luminosity and $E_{\rm p}$ is the peak energy of the $\nu f_\nu$ spectrum, suggesting that at least some GRBs have a dominant emission component above the photosphere in the optically thin region. Since the SSC mechanism suffers some other criticisms (e.g., Piran et al. 2009; Resmi \& Zhang 2012), synchrotron radiation is a natural candidate (see also Veres et al. 2012). Because both a quasi-thermal photosphere emission component and a non-thermal component are predicted to co-exist in the standard fireball shock model (M\'esz\'aros \& Rees 2000; Pe'er et al. 2006), the non-detection of the photosphere component (or in some cases a weak component) would point towards a magnetically dominated jet in GRBs (Zhang \& Pe'er 2009; Daigne \& Mochkovitch 2002; Zhang et al. 2011). Such a scenario is favored in the Fermi era, although further independent support is needed to make a stronger case.

It has long been speculated that the physics in different BH jet systems may be essentially the same (Mirabel 2004; Zhang 2007). A comparative study between the properties of AGNs and GRBs may shed light on the nature of these jets in different scales. By comparing the radio-to-optical spectral properties of optically bright GRB afterglows and blazars, Wang \& Wei (2011) suggested that GRB afterglows share a similar emission process with high frequency-peaked BL Lac objects. Wu et al. (2011) found that GRBs and blazars may have a similar intrinsic synchrotron luminosity in the co-moving frame, which is Doppler-boosted by different degrees in different systems, with GRBs having a higher Doopler boosting factor ($\delta$) than blazars. Lately, Nemmen et al. (2012) found that the energy dissipation efficiency of BH powered jets is similar over ten orders of magnitude in jet power, establishing a physical connection between AGNs and GRBs.

This letter compares the synchrotron radiation of AGNs with the prompt gamma-rays of GRBs and study their radiation efficiency ($\varepsilon$) and jet composition of AGNs and GRBs. We make use of our detailed modeling results for blazars and NLS1s to infer more physical parameters. The data and sample are presented in \S 2. A comparison of GRBs and AGNs in the jet power-luminosity plane is presented in \S 3. Jet radiation efficiency and  magnetization degree ($\sigma$) are presented in \S 4. Conclusions are presented in \S 5.

\section{Samples and Data}
\label{sect:data}
\begin{figure}
\includegraphics[width=3.0in,height=3.3in]{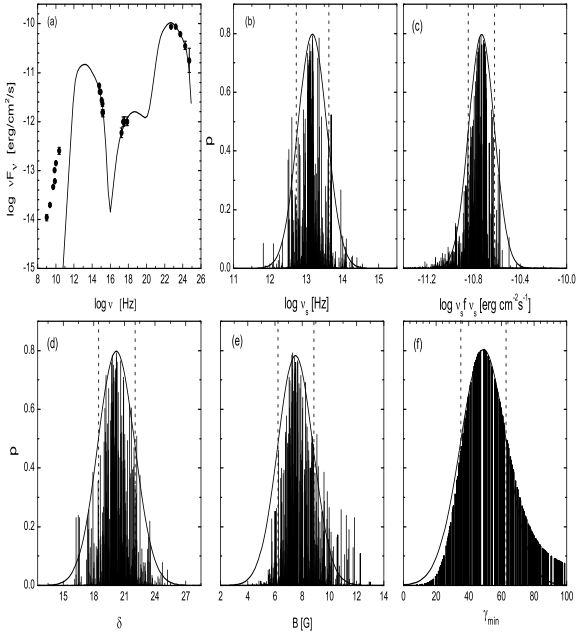}
\caption{An example (PKS 1454-354) of our SED fits and the probability distributions of the parameters along with our Gaussian fits.
The vertical lines mark the $1\sigma$ ranges of the parameters.}\label{Fig:SED_fit}
\end{figure}

Our AGN sample includes 19 BL Lacs, 23 FSRQs, and four NLS1s. They are GeV-TeV sources and 51 well-sampled broad-band SEDs are available for these sources. The SEDs are taken from Zhang et al. (2012b)\footnote{Thirty-four SEDs from 24 BL Lacs are reported in Zhang et al. (2012b). Twenty-four SEDs from 19 BL Lacs yield constraints on model parameters in 1$\sigma$ significance level. We include only these SEDs in this analysis.} and Abdo et al. (2009a, 2010). We fit the SEDs with a synchrotron+SSC model for BL Lacs. The details of our SED fits for the BL Lacs can be found in Zhang et al. (2012b). For NLS1s and FSRQs, a synchrotron+SSC+EC model is used to fit the SEDs. The external photon field of the EC process is taken as emission from the broad-line regions (BLRs) of FSRQs and NLS1s. The radiation from the BLRs is taken as a black body spectrum, with energy density measured in the comoving frame as $U_{\rm BLR}^{'}=3.76\times10^{-2}\Gamma^2$ erg/cm$^3$ (Ghisellini et al. 2010), where $\Gamma$ is the bulk Lorentz factor of the radiating region. Most AGNs in our sample are blazars. Since for blazars we are likely looking at the jet within the $1/\Gamma$ cone, and that the probability is the highest at the rim of the cone, we take $\delta \sim \Gamma$ in all the calculations, where $\delta$ is the beaming factor. The radiation region is taken as a homogeneous sphere with radius $R$, which is given by $R=c\delta \Delta t$. The minimum variability timescale is taken as $\Delta t=12$ hr for the FSRQs and NLS1s. The electron distribution is taken as a broken power law, which is characterized by a normalization term ($N_0$), a break energy $\gamma_{\rm b}$ and indices ($p_1$ and $p_2$) in the range $[\gamma_{\min}, \gamma_{\max}]$. These parameters are preliminarily derived from the observed SEDs and are further refined in our SED modeling (Zhang et al. 2012b). $N_{0}$ and $\gamma_{\rm b}$ depend on the synchrotron peak frequency ($\nu_{\rm s}$) and peak flux ($\nu_{\rm s} f_{\nu_{\rm s}}$) as well as other model parameters. We let $\nu_{\rm s}$ and $\nu_{\rm s} f_{\nu_{\rm s}}$ as free parameters in stead of $N_0$ and $\gamma_{\rm b}$. $\gamma_{\max}$ is poorly constrained, but it does not significantly affect our results. We fix it at a large value. Therefore, the free parameter set of our SED modeling is $\{B$, $\delta, \nu_{\rm s}, \nu_{\rm s} f_{\nu_{\rm s}}, \gamma_{\min}\}$, where $B$ is the magnetic field strength of the radiating region. We randomly generate a parameter set in broad spaces and measure the consistency between the model result and data with a probability $p\propto e^{-\chi^2_{\rm r}/2}$, where $\chi^2_{\rm r}$ is the reduced $\chi^2$. The center values and $1\sigma$ uncertainties of these parameters are derived from Gaussian fits to the profiles of the $p$ distributions. Since we set $\gamma_{\min}\geqslant 2$, the $p$ distributions of $\gamma_{\min}$ for some sources have a cutoff at $\gamma_{\min}=2$. The $\gamma_{\min}$ of NLS1s are poorly constrained in $1\sigma$ confidence level. Taking PKS 1454-354 as an example, Figure \ref{Fig:SED_fit} shows our best SED fits and the probability distributions of the parameters. Our results for the FSRQs and NLS1s are reported in Table 1. The observed SEDs with our best model fitting parameters are presented online\footnote{http://xil.bao.ac.cn/online.pdf}.

We adopt the conventional assumption that the jet power is carried by electrons and protons with one-to-one ratio, magnetic fields, and radiation, i.e., $P_{\rm jet} = \sum_i P_{\rm i}$, where $P_{\rm i}=\pi R^2 \Gamma^2 c U^{'}_{\rm i}$ and $U^{'}_{\rm i} (i={\rm e,\ p,\ B, \ r})$ is the energy density associated with each ingredient measured in the co-moving frame (Ghisellini et al. 2010). Note that the radiation power $P_{\rm r}$ should be a part of $P_{\rm jet}$ before radiation, since $P_{\rm e}$ is only the power carried by the electrons after radiation. It is estimated with the bolometric luminosity $L_{\rm bol}$, i.e., $P_{\rm r}=\pi R^2\Gamma^2 cU_{\rm r}^{'}=L_{\rm obs}\Gamma^2/4\delta{^4}\approx L_{\rm bol}/4\delta{^2}$. The radiation efficiency and the magnetization parameter of the jets can be calculated by
\begin{equation}
\varepsilon=P_{\rm r}/P_{\rm jet},
\end{equation}
\begin{equation}
\sigma=P_{\rm B}/(P_{\rm p}+P_{\rm e}+P_{\rm r}),
\end{equation}
respectively. The jet luminosities (bolometric $L_{\rm bol, jet}$, synchrotron peak $L_{\rm s, jet}$, and IC peak $L_{\rm c, jet}$) are derived with $L_{\rm w, jet}=f_{\rm b}L_{\rm w, iso}$, where ``w'' stands for ``bol'', ``s'', or ``c'', and $f_{\rm b}=1-\cos\theta\approx\theta^{2}/2\approx1/2\Gamma^{2}$ is the relativistic beaming correction factor. Our results are reported in Table 2.

The GRB sample includes 52 typical GRBs reported by Nemmen et al. (2012), excluding two low-luminosity GRBs (980425 and 060218) since they may have a different origin from most high-luminosity long GRBs (Liang et al. 2007). Their isotropic gamma-ray luminosity, jet luminosity, and jet power are calculated with $L_{\rm iso}=E_{\gamma, \rm iso}(1+z)/T_{90}$, $L_{\rm jet}=f_{\rm b} L_{\rm iso}$, and $P_{\rm jet}=f_{\rm b}(E_{\rm k, iso}+E_{\gamma, \rm iso})(1+z)/T_{90}$, respectively, where $T_{90}$ is the duration and $E_{\gamma, \rm iso}$ is the equivalent isotropic gamma-ray energy of GRBs, $f_{\rm b}$ is the beaming factor estimated with the jet opening angles derived from the observed break in their afterglow lightcurves, and $E_{\rm k, iso}$ is the isotropic kinetic energy of GRB fireballs estimated from the X-ray luminosity during the afterglow phase using the standard afterglow model. The GRB radiation efficiency is defined as $\varepsilon=E_{\gamma, \rm iso}/(E_{\gamma, \rm iso}+E_{\rm k, iso})$ (Zhang et al. 2007).

\section{GRB and AGN Sequence in the Luminosity - Jet Power Plane}
We first plot $P_{\rm jet}$ as a function of $L_{\rm s,iso}$, $L_{\rm s, jet}$, $L_{\rm bol, jet}$, and $L_{\rm c, jet}$ for AGNs in Figure 2. We make the best linear fits to the data with the minimum $\chi^2$ technique by considering the errors in both X and Y axes, if available, for the FSRQs and BL Lacs, respectively. The derived $P_{\rm jet}-L$ relations are marked in Figure 2\footnote{A summary of our fitting results and correlation analysis is available in the online materials, http://xil.bao.ac.cn/online.pdf.}. Tight correlations between $P_{\rm jet}$ and luminosities are observed for FSRQs, but not for BL Lacs. We do not make correlation analysis for NLS1s because we have only four NLS1s in our sample. Since the prompt emission mechanism of GRBs is not identified, we add the GRBs to these $L-P_{\rm jet}$ planes by plotting their $P_{\rm jet}$ against $L_{\rm iso}$ or $L_{\rm jet}$. Our best linear fit results are shown in Figure 2. Interestingly, the $L_{\rm s, jet}-P_{\rm jet}$ relation of FSRQs is well consistent with the $L_{\rm jet}-P_{\rm jet}$ relation of GRBs. We have $\log P_{\rm jet}=(14.2\pm3.6)+(0.72\pm0.08)\log L_{\rm s, jet}$ with $r=0.75$ and $p=4.3\times10^{-5}$ for the FSRQs, and $\log P_{\rm jet}=(14.1\pm2.8)+ (0.73\pm 0.06)\log L_{\rm jet}$ with $r=0.84$ and $p=4.7\times10^{-15}$ for the GRBs. We make best linear fits to the combined FSRQ and GRB sample and obtain $P_{\rm jet}\propto L_{\rm s, jet}^{0.79\pm0.01}$, with a Pearson correlation coefficient of $r=0.98$ (chance probability  $p\sim0$) and a dispersion of 0.44 dex. NLS1s are roughly in the $3\sigma$ confidence band of this relation, but most BL Lacs are out of this band. This relation spans ten orders of magnitude in luminosity with a small dispersion, indicating that both FSRQs and GRBs form a well sequence. Although tight $L_{\rm s, iso}-P_{\rm jet}$, $L_{\rm c, jet}-P_{\rm jet}$ and $L_{\rm bol, jet}-P_{\rm jet}$ relations are also found for the FSRQs, their intercepts are significantly different from the  $L_{\rm jet}-P_{\rm jet}$ relation of the GRBs. These results suggest that the physical properties between FSRQs and GRBs are likely similar and the dominant radiation mechanism of GRBs may be the synchrotron radiation of relativistic electrons in the jets.

\section{Radiation Efficiency and Jet Composition}
The standard internal shock model of GRBs predicts $\varepsilon\sim 5\%$ (e.g., Kumar 1999; Panaitescu et al. 1999). Some GRBs satisfy such a constraint, but most of them do not(Zhang et al. 2007; see also Fan \& Piran 2006). Dissipative photosphere emission (Lazzati et al. 2011) and internal-collision-induced magnetic reconnection and turbulence (ICMART) in a Poynting-flux-dominated wind (Zhang \& Yan 2011) have been suggested to achieve high radiation powers of GRBs. These two scenarios invoke distinctly different jet composition. While the photosphere model invokes a hot, matter-dominated fireball, the ICMART model invokes a magnetically dominated emitter in the emission region.

Figure 3(a) shows $\varepsilon$ as a function of $L_{\rm bol, jet}$ for AGNs and $\varepsilon$ as a function of $L_{\rm jet}$ for GRBs. The $\varepsilon$ of both GRB and FSRQ jets are comparable, with a large fraction being greater than 10\%, and they increase with $L_{\rm bol, jet}$ or $L_{\rm jet}$ with similar power-indices. Lloyd-Ronning \& Zhang (2004) reported a similar correlation between $\varepsilon$ and $E_{\gamma, \rm iso}$ for GRBs. Our best linear fits yield $\varepsilon\propto L_{\rm jet}^{0.41\pm0.05}$ with $r=0.4$ and $p=0.003$ for the GRBs and $\varepsilon\propto L_{\rm bol, jet}^{0.35\pm0.05}$ with $r=0.78$ and $p=1.1\times10^{-5}$ for the FSRQs. The slopes of both FSRQs and GRBs are consistent within error bars. The BL Lacs have a lower efficiency (normally $0.03\%\sim 10\%$) than FSRQs and GRBs. A large fraction of the BL Lacs are out of the 3$\sigma$ confidence band of the $\varepsilon-L_{\rm bol,jet}$ relation for the FSRQs. A weak $L_{\rm bol, jet}-\varepsilon$ correlation is also found for the BL Lacs alone, which is $\varepsilon\propto L_{\rm bol, jet}^{0.82\pm0.10}$ with $r=0.5$ and $p=0.013$. The index is different from that of the FSRQs and GRBs.

We further study the jet composition of AGNs with our modeling results. Even though current GRB modeling does not allow us to constrain the magnetization parameter $\sigma$, this can be done for the AGN sample. In Figure 3(b), we plot $\varepsilon$ against $\sigma$ for all AGNs in our sample. The FSRQs tend to have higher $\varepsilon$ and $\sigma$ values than the BL Lacs. Their $\sigma$ values are close to or exceeding unity. An anti-correlation between $\varepsilon$ and $\sigma$ is found for the FSRQs, i.e., $\varepsilon\propto \sigma^{-0.32\pm 0.09}$ with $r=-0.63$ and $p=0.001$. This is due to $P_{\rm B}$ dominates $P_{\rm jet}$ for FSRQs, and an anti-correlation between $\varepsilon$ and $\sigma$ may be expected from Eqs (1) and (2). NLS1s show similar trend, but systematically have a lower $\varepsilon$ than FSRQs. BL Lacs have lower $\varepsilon$ and $\sigma$ values, and a weak correlation is found, i.e., $\varepsilon\propto \sigma^{0.57\pm 0.09}$ with $r=0.55$ and $p=0.005$. The dramatic difference of the $\varepsilon-\sigma$ correlations between FSRQs and BL Lacs may further signal the different jet properties of two kinds of sources. The FSRQ jets are likely highly magnetized and the BL Lac jets are less radiation efficiency and matter dominated. Since GRBs have a similar radiation efficiency and efficiency-luminosity dependence as FSRQs, one may suggest that the jet properties of GRBs are analogous to FSRQs. This supports the idea that GRB emission is due to magnetic dissipation in a highly magnetized jet (e.g. Zhang \& Yan 2011).

\section{Conclusions and Discussion}
We have presented a comparative study of GRBs and radio loud AGNs, making use of our systematical SED modeling results for a GeV-TeV selected sample of AGNs. We show that the $P_{\rm jet}-L_{s, \rm jet}$ relation of FSRQs is consistent with the $P_{\rm jet}-L_{\rm jet}$ relation of GRBs. The radiation efficiencies of both FSRQs and GRBs are comparable and even increase with $L_{\rm bol, jet}$ with a similar power-law index. BL Lacs typically have a lower $\varepsilon$ and $L_{\rm bol, jet}$ than FSRQs, and a tentative $L_{\rm bol, jet}-\varepsilon$ relation is found with a different slope from that of the FSRQs. An anti-correlation between $\varepsilon$ and $\sigma$ is found for FSRQs, but this correlation is positive for the BL Lacs. Based on the analogy between GRBs and FSRQs, we suggest that GRBs are likely produced by synchrotron process in a magnetized jet with high radiation efficiency. The jets of NLS1s potentially share similar properties with FSRQs. The jets of BL Lacs, on the other hand, are low radiation efficiency and likely matter dominated.

Compared with AGNs, GRBs are less understood. There is a list of open questions in GRB physics, including jet composition, energy dissipation and radiation mechanisms (e.g., Zhang 2011). Our comparative study between GRBs and AGNs shed new light on some of these open questions of GRBs. For example, the clear jet power - luminosity correlation suggests that the dominant radiation mechanism of GRB prompt emission is similar to the low energy peak of SEDs for blazars, namely, synchrotron radiation (see also an independent study of Uhm \& Zhang 2013). The close analogy between GRBs and FSRQs in their radiation efficiency - luminosity dependence and the fact that FSRQs have a moderate to high magnetization parameter $\sigma$ suggest that GRB emission is likely from energy dissipation in a highly magnetized jet (e.g., Zhang \& Yan 2011). The high radiation efficiency argument alone may not disfavor the photosphere model of GRBs. However, when combining the luminosity - jet power correlation as presented in Figure 2, the magnetic dissipation model is further favored since it invokes synchrotron radiation as the dominant radiation mechanism. All these are also consistent with Roming et al. (2006), who discovered that some GRBs with high radiation efficiency tend to have tight early UVOT upper limits, which could be caused by suppression of the reverse shock emission in a magnetized jet (e.g., Zhang \& Kobayashi 2005; Mimica et al. 2009).

Above conclusions are based on the leptonic model of radiation for the AGNs and comparative analysis between the AGNs and GRBs. The derivation of $P_{\rm jet}$ of AGNs is essential for our analysis. The estimate of $P_{\rm e}$ and $P_{\rm p}$, especially $P_{\rm p}$ are significantly affected by the $\gamma_{\rm min}$ values, which are simply taken as unity in previous work (e.g., Ghisellini et al. 2010). As shown in Table 1, we find that the typical $\gamma_{\min}$ value is 45, which lowers both $P_{\rm e}$ and $P_{\rm p}$, with more drastic decrease of $P_{\rm p}$, as compared with Ghisellini et al. (2010). As a result, both $\varepsilon$ and $\sigma$ derived in our paper are systematically higher than those derived in Ghisellini et al. (2010). Note that we assumed one cool proton for one relativistic electron in jet power calculations. In case of that the jet power is carried by positron-electron pairs, magnetic field, and radiation, but no protons, the main results for FSRQs hold and the conclusions of our analysis are still valid, but no significant luminosity-radiation efficiency and magnetization parameter-radiation efficiency correlations are found for BL Lacs.

\acknowledgments
We thank helpful discussion with G. Ghisellini, Jian-Yan Wei, Xue-Feng Wu, and Zi-Gao Dai. This work is supported by the National Basic Research Program (973 Programme) of China (Grant 2009CB824800), the National Natural Science Foundation of China (Grants 11078008, 11025313, 11133002, 10725313), Guangxi Science Foundation (2013GXNSFFA019001, 2011GXNSFB018063, 2010GXNSFC013011). BZ acknowledges support from NSF (AST-0908362).


\begin{deluxetable}{lccccccccc}
\tabletypesize{\footnotesize} \tablecolumns{10}\tablewidth{40pc} \tablecaption{Model parameters of SED fits for the FSRQs and NLS1s in our sample}\tablenum{1} \tablehead{\colhead{Source}&\colhead{redshift}&\colhead{$p_{1}$}
&\colhead{$p_{2}$}&\colhead{$\gamma_{\rm min}$}&\colhead{$\gamma_{\rm max}$}&\colhead{$\gamma_{\rm b}$}&\colhead{$N_{0}$}&\colhead{$\delta$}& \colhead{$B$[G]}}
\startdata
FSRQs&&&&&&&&&\\
\hline
3C 279&0.536&1.6&3.6&2$^{+20}_{-0}$&5000&219$\pm$37&(8.7$\pm$3.4)E3&12.0$\pm$0.5&5.9$\pm$0.3\\
3C 273&0.158&1.4&4.0&6$^{+19}_{-4}$&2000&328$\pm$79&(5.6$\pm$5.3)E3&7.4$\pm$0.9&8.5$\pm$1.6\\
3C 454.3&0.859&1.4&3.7&33$\pm$14&1700&203$\pm$25&(4.8$\pm$1.8)E3&17.6$\pm$0.6&6.6$\pm$0.6\\
PKS 1454-354&1.424&2.4&3.6&49$\pm$14&2000&276$\pm$99&(3.2$\pm$2.5)E5&20.2$\pm$1.8&7.5$\pm$1.3\\
PKS 0208-512&1.003&1.3&3.6&48$\pm$10&3000&227$\pm$82&(1.9$\pm$1.6)E3&15.2$\pm$1.3&5.9$\pm$1.1\\
PKS 0454-234&1.003&1.4&3.6&17$^{+18}_{-15}$&2500&184$\pm$56&(6.3$\pm$5.4)E2&20.0$\pm$1.9&6.6$\pm$0.8\\
PKS 0727-11&1.589&1.8&3.9&50$\pm$12&5000&254$\pm$61&(1.6$\pm$1.1)E4&20.6$\pm$1.2&5.4$\pm$1.1\\
PKS 0528+134&2.07&1.04&4.2&80$\pm$8&7000&222$\pm$54&(2.2$\pm$1.7)E3&18.4$\pm$1.3&5.9$\pm$1.2\\
4C 66.20 &0.657&1.8&3.2&48$\pm$11&3000&240$\pm$95&(1.7$\pm$1.6)E4&12.2$\pm$1.2&7.2$\pm$1.4\\
4C 29.45&0.729&2.0&4.3&49$\pm$8&5000&400$\pm$112&(5.6$\pm$4.7)E4&11.6$\pm$1.0&8.3$\pm$1.6\\
B2 1520+31&1.487&2.2&4.2&48$\pm$16&2500&283$\pm$93&(6.3$\pm$4.8)E4&20.8$\pm$1.6&4.3$\pm$0.9\\
PKS 0420-01&0.916&2.64&4.06&51$\pm$11&5000&385$\pm$118&(2.5$\pm$1.4)E6&12.8$\pm$0.7&8.2$\pm$1.1\\
1Jy 1308+326&0.997&2.2&3.4&44$\pm$10&8000&353$\pm$129&(4.9$\pm$4.5)E5&12.6$\pm$0.9&3.4$\pm$0.9\\
PKS 1510-089&0.36&1.12&3.72&20$\pm$12&1500&305$\pm$32&(4.2$\pm$1.5)E2&11.0$\pm$0.5&3.1$\pm$0.5\\
4C 28.07&1.213&1.4&3.96&41$\pm$10&10000&223$\pm$51&(4.3$\pm$3.1)E3&14.6$\pm$1.1&6.9$\pm$1.0\\
PMN 2345-1555&0.621&2.0&4.0&46$\pm$11&8000&202$\pm$58&(1.9$\pm$1.6)E4&13.8$\pm$1.3&6.0$\pm$1.3\\
S3 2141+17 &0.213&1.8&3.9&53$\pm$10&13000&504$\pm$154&(7.4$\pm$7.2)E3&8.0$\pm$1.0&12.8$\pm$1.8\\
S4 0133+47&0.859&1.4&3.2&49$\pm$8&4000&191$\pm$65&(2.7$\pm$2.3)E3&13.1$\pm$1.2&10.5$\pm$1.5\\
S4 0917+44&2.19&1.3&3.54&10$^{+19}_{-8}$&5000&213$\pm$66&(3.2$\pm$2.9)E3&18.2$\pm$1.3&9.8$\pm$1.8\\
PKS 0227-369&2.115&1.88&4.1&86$\pm$8&2800&331$\pm$102&(1.0$\pm$0.6)E5&17.8$\pm$1.0&5.9$\pm$0.9\\
PKS 0347-211&2.944&1.2&3.9&59$\pm$36&4000&222$\pm$68&(5.0$\pm$3.7)E2&26.2$\pm$1.5&10.0$\pm$1.5\\
PKS 2325+093&1.843&1.4&4.6&72$\pm$6&900&354$\pm$107&(4.4$\pm$3.6)E3&17.6$\pm$1.6&15.1$\pm$1.6\\
PKS 1502+106&1.839&1.2&3.6&14$^{+19}_{-12}$&3000&192$\pm$66&(2.6$\pm$2.5)E2&27.0$\pm$2.3&7.1$\pm$1.5\\
\hline
NLS1s&&&&&&&&&\\
\hline
PKS 1502+036&0.409&1.2&4.5&55&10000&435$\pm$129&(2.2$\pm$3.8)E2&9.0$\pm$2.0&6.6$\pm$2.4\\
PKS 2004-447&0.24&1.2&4.3&30&3000&311$\pm$42&(1.1$\pm$0.5)E3&6.0$\pm$0.4&7.2$\pm$1.4\\
PMN J0948+0022&0.5846&1.2&4.1&18&35000&297$\pm$40&(2.0$\pm$1.0)E3&9.8$\pm$0.6&4.8$\pm$0.6\\
1H 0323+342&0.061&1.2&4.0&11&25000&637$\pm$128&(1.4$\pm$0.9)E3&3.8$\pm$0.3&8.4$\pm$1.5\\
\enddata
\end{deluxetable}

\begin{deluxetable}{lcccccc}
\tabletypesize{\footnotesize}
\tablecolumns{7}\tablewidth{38pc} \tablecaption{The data of the AGNs in our sample}\tablenum{2} \tablehead{\colhead{Source\tablenotemark{a}}  & \colhead{$\log L_{\rm s}$}  & \colhead{$\log L_{\rm c}$}
& \colhead{$\log L_{\rm bol}$} &  \colhead{$\log P_{\rm jet}$} & \colhead{$\varepsilon$} & \colhead{$\sigma$}\\
\colhead{}& \colhead{(erg/s)} & \colhead{(erg/s)} & \colhead{(erg/s)}&  \colhead{(erg/s)} & \colhead{} &
\colhead{}}

\startdata
BL Lacs\tablenotemark{a}&&&&&&\\
\hline
Mkn 421$^{\rm L}$&44.87&44.31&45.80$\pm$0.01&44.39$^{+0.37}_{-0.29}$&(7.7$^{+10.0}_{-7.7}$)E-3&(2.2$^{+11.0}_{-0.7}$)E-2\\
Mkn 501$^{\rm L}$&44.29&43.77&45.25$\pm$0.02&43.94$^{+0.95}_{-0.20}$&(2.6$^{+6.0}_{-2.2}$)E-2&(7.2$^{+9.2}_{-5.3})$E-2\\
W Com$^{\rm L}$&44.91&44.86&46.00$\pm$0.05&44.65$^{+0.34}_{-0.15}$&(2.5$^{+2.0}_{-0.9}$)E-2&(1.9$^{+1.9}_{-1.1}$)E-2\\
BL Lacertae$^{\rm L}$&45.06&44.79&46.02$\pm$0.07&45.21$^{+0.40}_{-0.15}$&(4.5$^{+4.2}_{-1.8}$)E-3&(7.0$^{+7.7}_{-4.1}$)E-3\\
1ES 1959+650$^{\rm L}$&44.95&43.92&45.82$\pm$0.03&45.75$^{+0.02}_{-0.13}$&(2.4$^{+0.8}_{-1.1}$)E-3&(1.3$\pm$0.4)E-2\\
1ES 2344+514$^{\rm L}$&44.10&43.71&44.97$\pm$0.40&44.55$^{+0.02}_{-0.09}$&(3.9$^{+5.1}_{-3.9}$)E-3&(6.8$\pm$0.2)E-3\\
PKS 2155-304$^{\rm L}$&45.87&45.35&46.79$\pm$0.01&44.61$^{+0.33}_{-0.18}$&(1.5$^{+1.6}_{-1.3}$)E-2&(5.9$^{+8.0}_{-3.1}$)E-2\\
1ES 1101-232$^{\rm L}$&45.54&44.60&46.35$\pm$0.02&44.61$^{+0.18}_{-0.12}$&(9.7$^{+4.3}_{-3.1}$)E-2&0.34$^{+0.36}_{-0.08}$\\
Mkn 501$^{\rm H}$&45.46&44.96&46.29$\pm$0.12&44.95$^{+2.92}_{-0.23}$&(2.5$^{+17.0}_{-1.9}$)E-2&(3.7$^{+4.8}_{-3.4}$)E-4\\
W Com$^{\rm H}$&45.00&45.37&46.44$\pm$0.08&44.88$^{+0.63}_{-0.04}$&(4.6$^{+6.7}_{-1.0}$)E-2&(7.6$^{+1.7}_{-4.3}$)E-3\\
BL Lacertae$^{\rm H}$&44.69&44.81&45.96$\pm$0.04&45.79$^{+0.21}_{-0.17}$&(9.3$^{+6.6}_{-6.1}$)E-4&(3.6$^{+2.4}_{-1.0}$)E-4\\
PKS 2005-489$^{\rm H}$&45.34&44.11&46.38$\pm$0.01&44.99$^{+0.62}_{-0.18}$&(3.5$^{+5.6}_{-2.9}$)E-3&0.11$\pm0.07$\\
1ES 1959+650$^{\rm H}$&45.26&44.83&46.33$\pm$0.02&46.89$^{+2.52}_{-0.17}$&(4.8$^{+28.0}_{-2.2}$)E-4&(6.7$^{+4.1}_{-5.2}$)E-5\\
PKS 2155-304$^{\rm H}$&46.12&46.83&47.82$\pm$0.04&45.44$^{+0.19}_{-0.10}$&(8.9$^{+4.1}_{-2.3}$)E-2&(3.8$^{+1.8}_{-0.7}$)E-3\\
3C 66A&46.60&47.14&48.09$\pm$0.05&45.77$^{+0.10}_{-0.06}$&(9.1$^{+4.1}_{-3.3}$)E-2&(6.8$^{+1.1}_{-1.6}$)E-3\\
PG1553+113&46.61&46.11&47.50$\pm$0.03&44.91$^{+0.04}_{-0.04}$&(9.5$\pm3.7$)E-2&0.49$^{+0.13}_{-0.03}$\\
1ES 1218+30.4&45.39&45.19&46.36$\pm$0.01&44.14$^{+0.47}_{-0.17}$&0.10$^{+0.15}_{-0.10}$&0.11$^{+0.26}_{-0.08}$\\
1ES 1011+496&46.20&45.64&47.05$\pm$0.03&46.07$^{+1.44}_{-0.19}$&(1.4$^{+4.7}_{-0.7}$)E-2&(2.1$^{+2.3}_{-1.6}$)E-2\\
PKS 1424+240&47.05&46.39&47.82$\pm$0.01&45.31$^{+0.12}_{-0.08}$&(7.5$^{+3.5}_{-3.0}$)E-2&0.53$^{+0.43}_{-0.02}$\\
1ES 0806+524&44.87&44.35&45.84$\pm$0.01&46.23$^{+0.28}_{-0.31}$&(7.0$^{+6.5}_{-6.8}$)E-4&(2.4$^{+5.6}_{-0.7}$)E-3\\
Mkn 180&43.88&43.63&44.82$\pm$0.05&44.33$^{+0.10}_{-0.16}$&(2.2$^{+0.7}_{-1.0}$)E-2&(2.3$\pm$1.9)E-2\\
RGB J0152+017&43.89&43.93&45.07$\pm$0.07&45.45$^{+2.41}_{-0.01}$&(4.3$^{+24.0}_{-0.7}$)E-3&(3.8$^{+0.01}_{-3.0}$)E-4\\
H1426+428&44.74&45.27&46.30$\pm$0.04&44.91$^{+0.85}_{-0.13}$&(8.4$^{+16.0}_{-2.7}$)E-2&(1.3$^{+1.0}_{-0.8}$)E-3\\
PKS 0548-322&44.31&43.54&45.20$\pm$0.02&44.98$^{+2.72}_{-0.03}$&(1.2$^{+7.2}_{-0.2}$)E-2&(8.2$^{+0.5}_{-6.8}$)E-3\\
\hline
FSRQs&&&&&&\\
\hline
3C 279&46.44$\pm$0.05&46.92$\pm$0.05&47.80$\pm$0.01&45.93$\pm$0.17&0.13$\pm$0.05&0.31$\pm$0.06\\
3C 273&46.25$\pm$0.13&45.95$\pm$0.30&47.11$\pm$0.03&45.58$\pm$0.18&0.15$\pm$0.07&0.42$\pm$0.25\\
3C 454.3&47.60$\pm$0.10&48.30$\pm$0.05&49.06$\pm$0.01&46.47$\pm$0.08&0.32$\pm$0.06&0.35$\pm$0.08\\
PKS 1454-354&47.35$\pm$0.11&48.13$\pm$0.25&48.92$\pm$0.04&46.34$\pm$0.14&0.23$\pm$0.09&0.85$\pm$0.44\\
PKS 0208-512&46.87$\pm$0.13&47.54$\pm$0.15&48.34$\pm$0.02&45.92$\pm$0.12&0.29$\pm$0.10&0.53$\pm$0.26\\
PKS 0454-234&46.94$\pm$0.11&47.77$\pm$0.09&48.53$\pm$0.02&46.22$\pm$0.14&0.13$\pm$0.05&2.26$\pm$0.95\\
PKS 0727-11&47.18$\pm$0.15&48.19$\pm$0.15&48.91$\pm$0.03&46.16$\pm$0.11&0.33$\pm$0.09&0.53$\pm$0.23\\
PKS 0528+134&47.54$\pm$0.14&48.38$\pm$0.25&48.99$\pm$0.06&46.24$\pm$0.11&0.42$\pm$0.13&0.17$\pm$0.09\\
4C 66.20 &46.41$\pm$0.15&46.82$\pm$0.12&47.77$\pm$0.02&45.72$\pm$0.15&0.19$\pm$0.07&0.96$\pm$0.53\\
4C 29.45&46.59$\pm$0.17&46.72$\pm$0.22&47.60$\pm$0.03&45.69$\pm$0.15&0.15$\pm$0.06&1.09$\pm$0.58\\
B2 1520+31&46.73$\pm$0.13&47.95$\pm$0.25&48.64$\pm$0.05&45.98$\pm$0.14&0.26$\pm$0.10&0.57$\pm$0.30\\
PKS 0420-01&46.77$\pm$0.13&47.06$\pm$0.13&48.01$\pm$0.02&45.98$\pm$0.13&0.16$\pm$0.05&0.47$\pm$0.16\\
1Jy 1308+326&46.24$\pm$0.20&47.33$\pm$0.15&48.20$\pm$0.02&45.99$\pm$0.24&0.26$\pm$0.15&0.05$\pm$0.03\\
PKS 1510-089&45.86$\pm$0.05&46.79$\pm$0.14&47.46$\pm$0.03&45.42$\pm$0.09&0.23$\pm$0.05&0.24$\pm$0.08\\
4C 28.07&46.84$\pm$0.17&47.30$\pm$0.14&48.09$\pm$0.02&45.87$\pm$0.13&0.20$\pm$0.06&0.59$\pm$0.25\\
PMN 2345-1555&46.13$\pm$0.13&46.70$\pm$0.20&47.46$\pm$0.04&45.66$\pm$0.19&0.08$\pm$0.04&1.97$\pm$1.18\\
S3 2141+17 &46.03$\pm$0.11&45.47$\pm$0.15&46.78$\pm$0.02&45.51$\pm$0.21&0.07$\pm$0.04&5.81$\pm$3.42\\
S4 0133+47&46.8$\pm$0.13&46.94$\pm$0.13&48.00$\pm$0.01&45.95$\pm$0.14&0.17$\pm$0.06&1.86$\pm$0.87\\
S4 0917+44&47.46$\pm$0.15&47.86$\pm$0.13&48.77$\pm$0.02&46.27$\pm$0.15&0.24$\pm$0.09&0.61$\pm$0.25\\
PKS 0227-369&47.31$\pm$0.13&48.08$\pm$0.16&48.80$\pm$0.03&46.08$\pm$0.09&0.42$\pm$0.10&0.24$\pm$0.09\\
PKS 0347-211&47.75$\pm$0.15&48.42$\pm$0.13&49.13$\pm$0.03&46.42$\pm$0.12&0.19$\pm$0.06&2.61$\pm$0.99\\
PKS 2325+093&48.10$\pm$0.13&47.98$\pm$0.17&48.84$\pm$0.03&46.41$\pm$0.13&0.22$\pm$0.08&1.92$\pm$0.88\\
PKS 1502+106&47.60$\pm$0.14&48.60$\pm$0.22&49.36$\pm$0.04&46.55$\pm$0.15&0.22$\pm$0.09&1.81$\pm$0.94\\
\hline
NLS1s&&&&&&\\
\hline
PKS 1502+036&45.37$\pm$0.20&45.35$\pm$0.50&46.20$\pm$0.08&45.02$\pm$0.44&0.05$\pm$0.05&4.54$\pm$5.57\\
PKS 2004-447&45.10$\pm$0.07&44.73$\pm$0.22&45.80$\pm$0.02&44.76$\pm$0.11&0.08$\pm$0.02&0.93$\pm$0.43\\
PMN J0948+0022&46.11$\pm$0.06&46.54$\pm$0.10&47.32$\pm$0.02&45.54$\pm$0.09&0.16$\pm$0.04&0.18$\pm$0.06\\
1H 0323+342&44.67$\pm$0.05&44.00$\pm$0.20&45.49$\pm$0.01&44.40$\pm$0.10&0.22$\pm$0.06&0.47$\pm$0.23\\
\enddata
\tablenotetext{a}{The data of BL Lacs are taken from Zhang et al. (2012b). The source names marked with ``H'' or ``L'' are for a high or low state of the sources as defined in Zhang et al. (2012b).}
\end{deluxetable}

\begin{figure}
\includegraphics[scale=0.3]{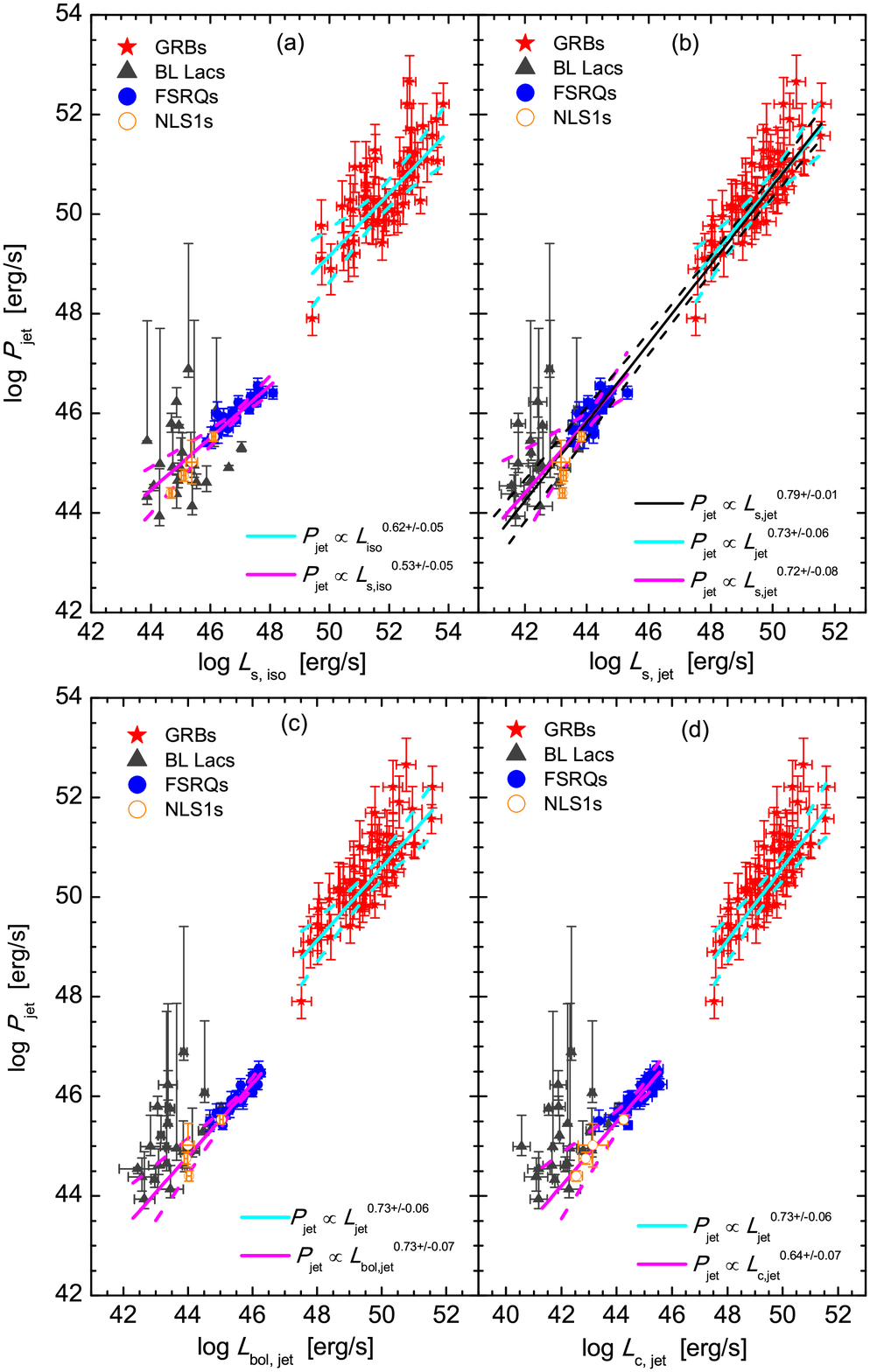}
\caption{Jet power as a function of 
(a) isotropic synchrotron radiation peak luminosity of AGNs and isotropic gamma-ray luminosity of GRBs, (b) geometrically-corrected synchrotron radiation peak luminosity of AGNs and GRB jet luminosity, (c) geometrically-corrected bolometric luminosity of AGNs and jet luminosity of GRBs, and (d) geometrically-corrected IC peak luminosity of AGNs and GRB jet luminosity. Solid and dashed lines are the best linear fits and their 3$\sigma$ confidence bands for the GRBs ({\em Cyan}) and FSRQs ({\em magenta}). In panel (b), the best fit and its 3$\sigma$ confidence band ({\em black lines}) to the combined GRB and FSRQ sample are also shown. }\label{Fig:PjetL}
\end{figure}

\begin{figure}
\includegraphics[scale=0.3]{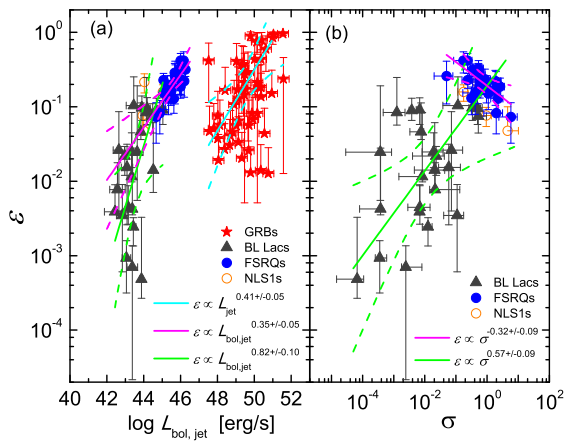}
\caption{Jet radiation efficiency ($\varepsilon$) as a function of the jet luminosity (geometrically-corrected bolometric luminosity of AGNs and prompt gamma-ray luminosity of GRBs) and the magnetization parameter ($\sigma$) of AGNs in our sample. Color lines are the best linear fits along with 3$\sigma$ confidence bands to the data of the GRBs ({\em cyan}), FSRQs ({\em magenta}), and BL Lacs ({\em green}).}\label{Fig:PrPjet}
\end{figure}

\label{lastpage}

\end{document}